\documentclass[preprint,showpacs,preprintnumbers,amsmath,amssymb]{revtex4}

\usepackage{graphicx}
\usepackage{dcolumn}
\usepackage{bm}

\begin{document}


\preprint{APS/123-QED}

\title{Isospin effects in a covariant transport approach to spallation reactions:
Analysis of $p+Fe$ and $Pb$ reactions at $0.8$, $1.2$ and $1.6$ GeV}

\author{Khaled Abdel-Waged\footnote{khelwagd@yahoo.com}}
\affiliation{Umm Al-Qura University, Faculty of Applied Science,
Physics Department, Makkah Unit 126, P.O. Box 7047, Saudi Arabia}
\author{Nuha Felemban}
\affiliation{Umm Al-Qura University, Faculty of Applied Science,
Physics Department, Makkah Unit 126, P.O. Box 7047, Saudi Arabia}
\affiliation{King Saud University, Faculty of Science,
Physics and Astronomy Department, Riyadh, Saudi Arabia}
\author{Theodoros Gaitanos}
\affiliation{Institut f\"{u}r Theoretische Physik,
Justus-Liebig-Universit\"{a}t Giessen,
D-35392 Giessen, Germany}
\author{Graziella Ferini}
\affiliation{Laboratori Nazionali del Sud INFN, I-95123 Catania,
Italy}
\author{Massimo Di Toro}
\affiliation{Laboratori Nazionali del Sud INFN, I-95123 Catania,
Italy}

\date{\today}

\begin{abstract}
We have investigated the influence of different non-linear relativistic mean field
models ($NL$, $NL\rho$  and $NL\rho\delta$) on spallation neutrons for p+Fe and Pb
reactions at 0.8, 1.2 and 1.6 GeV by means of a relativistic Boltzmann Uehling
Uhlenbeck (RBUU) approach plus a statistical multifragmentation (SM) decay model.
We find that the "evaporation shoulder", i.e. the neutron energy spectrum from
3 to 30 MeV, almost for any emission angle is quite sensitive to the isospin part
of the mean field.  For the more neutron-rich Pb-target the evaporation component
can be directly related to the low density behavior on the symmetry energy in the
thermal expansion phase of the excited compound system. It turns out that the
spallation data for the reactions under study are shown to be more consistent
with RBUU+SM employing  the $NL\rho$  effective lagrangian.
\end{abstract}
\pacs{25.40.Sc, 24.10.Lx, 25.40.Ep, 24.10.Pa}

\maketitle


\section{Introduction}

Spallation reactions are very important for their wide applications in
accelerator technology, production of energy, astrophysics and transmutation
of nuclear waste \cite{ref1,ref2,ref3,ref4,ref5,ref6}.
The theoretical description of  the spallation
process is essential in understanding better the physical mechanism of
each of the mentioned cases.

Several dynamical models have been constructed
for the theoretical description of spallation process
\cite{ref7,ref8,ref9,ref10,ref11}. All of them
have the same basis, they describe the reaction as a cascade of
nucleon-nucleon (NN) collisions, but employing different assumptions.
The main difference concerns implementing mean field dynamics. One can
distinguish the simplest models, which neglect features of the mean field
dynamics and employ constant static potential, like a class of Intra-nuclear
cascade (INC) models \cite{ref7}. Other, more sophisticated approaches comprise
changing field and real fluctuations obtained due to use of two and three
body potentials e.g., quantum molecular dynamics (QMD) models \cite{ref10}.
Boltzmann-Uehling-Uhlenbeck (BUU) type models use both non-relativistic
and relativistic mean field potentials \cite{ref9,ref12}. An important advantage
of relativistic mean field models is the distinct separation of scalar
and vector Lorentz-components of the nuclear mean field potential.
The scalar part is associated with the in-medium dependence of the effective
mass (Dirac mass), while the vector one changes the properties of the particle
momenta inside the hadronic environment. This separation between scalar and
vector self energies is very useful when extending mean field models to
asymmetric nuclear matter, in which the role of isovector mesons can be
clearly associated with the in-medium properties, symmetry potentials and
effective masses of protons and neutrons.
	
In this work, we study the double differential cross sections (DDCS) of
emitted neutrons in the reactions induced by 0.8, 1.2 and 1.6 GeV proton on targets
Fe and Pb by using a covariant transport model of a relativistic BUU (RBUU) type
\cite{ref12}.
The RBUU model is found to be a useful tool in describing the time evolution of
the nucleus-nucleus reaction dynamics \cite{ref13,ref14,ref15,ref16}.
As a genuine feature of transport
theories it has two important ingredients: a relativistic mean field (RMF) based
on quantum hadrodynamics \cite{ref17} and  isospin effects that account for the elastic
and inelastic channels in resonance production and decay (cf. Refs. \cite{ref13,ref18}
for different numerical realizations).  In our approach, the non-linear (NL)-RMF
mean-field
in RBUU consists of isoscalar and isovector parts with different Lorentz properties.
The isoscalar part is characterized by an attractive scalar ($\sigma$-meson) and a
repulsive vector field ($\omega$-meson), which is of importance in describing
saturation properties of nuclear matter \cite{ref17}. The isovector part of NL-RMF is also
characterized by a competition between vector ($\rho$-meson) and scalar ($\delta$-meson)
fields, which is responsible for the density dependence of the symmetry energy
\cite{ref16,ref19,ref20,ref21}.
It should be noted that, the scalar nature of the $\delta$-meson leads even to a new
interesting effect, important in the reaction dynamics, the splitting of the neutron
and proton effective masses in isospin asymmetric systems.  The numerical values of
the field parameters, fixed to nuclear matter and symmetry energy at saturation,
are chosen similar to the QMD parameters for a better comparison with the results
presented in \cite{ref10,ref22,ref23}.

An important feature of our transport model is that isospin effects are also
explicity included in the collision integral, see later. The results presented
here are based on the RBUU code presented in
Refs. \cite{ref13,ref14,ref16,ref20,ref21}, where also
the properties of the NL-RMF, cross sections and the collision integral are
discussed. For a meaningful comparison with spallation neutron data we use the
statistical multifragmenation (SM) model as an afterburner \cite{ref24}. 	

The paper is organized as follows. Section 2 defines the basic ingredients of the RBUU model.
In Sec.3, we apply the RBUU+SM code systematically to the experimental neutron DDCS
in the interactions of 0.8, 1.2 and 1.6 GeV proton on Fe and Pb at fixed angles
of $0^{o}$, $10^{o}$, $25^{o}$, $40^{o}$, $55^{o}$, $85^{o}$, $100^{o}$, $115^{o}$,
$130^{o}$, $145^{o}$ and $160^{o}$. We summarize and conclude this work in Sec. 4.

\section{Description of the RBUU model}

In this section, an outline of the RBUU approach is given, which is described
in detail in
Refs. \cite{ref13,ref14,ref16,ref20,ref21}. The RBUU equation has the form
\begin{eqnarray}
& & \left[
p^{*\mu} \partial_{\mu}^{x} + \left( p^{*}_{\nu} F^{\mu\nu}
+ m^{*} \partial_{x}^{\mu} m^{*}  \right)
\partial_{\mu}^{p^{*}}
\right] f(x,p^{*}) = {\cal I}_{\rm coll}
\label{rbuu}
\quad .
\end{eqnarray}
Eq.~\ref{rbuu} describes the evolution of the single particle distribution
function $f(x,p^{*})$ under the influence of a mean-field, which enters via
effective masses $m^{*}$, effective momenta $p^{*\mu}$ and the field tensor
$F^{\mu\nu} = \partial^\mu \Sigma^\nu -\partial^\nu \Sigma^\mu$, and $2$-body
collisions $I_{coll}$. $\Sigma_{s}$ and $\Sigma^{\mu}$ are the scalar and vector
self energies, respectively.

In actual simulations the test particle (relativistic Landau Vlasov (RLV))
method \cite{ref25,ref26} is used for the numerical treatment of  the Vlasov part, the
lhs of Eq.~(\ref{rbuu}). In this work we use a covariant Gaussian (in coordinate
and momentum space) shape for the test particles. In \cite{ref25} it was shown that the
use of a Gaussian shape for the test particles is appropriate to produce smooth
fields and it is possible to determine local quantities, such as densities,
currents, etc, without introducing additional grids.

With the introduced Gaussians the one-body phase space
distribution function $f(x,p^{*})$ is simulated in the following way
\begin{equation}
f(x,p^*) = \frac{1}{N_{test}} \sum_{i=1}^{A\cdot N_{test} }
             \int \limits_{-\infty}^{\infty} d\tau
             g \left( x- x_{i}(\tau )\right)
             g \left(p^{*} - p_{i}^{*}(\tau ) \right)
\label{fs_1}
\end{equation}
where $A$ is the number of nucleons and $N_{test}$ is the number of test
particles per nucleon. The four dimensional Gaussian weights of the test
particles take the form
\begin{equation}
g ( x- x_{i} (\tau ) ) =
\alpha_{s} \; \exp{(R_{i\mu}(x)R_{i}^{\mu}(x) /\sigma_{s}^{2})} \;
\delta \left[(x_\mu - x_{i\mu}(\tau )) u_{i}^{\mu}(\tau )  \right]
\label{gauss-x}
\end{equation}
and
\begin{equation}
g ( p^{*} - p_{i}^{*}(\tau ) ) = \alpha_{p} \;
\exp{(( p^{*} - p_{i}^{*} (\tau ) )^{2} / \sigma_{k}^{2})} \;
\delta \left[ p_{\mu}^{*} p_{i}^{*\mu}(\tau ) - m_{i}^{*2} \right]
\label{gauss-p}
\end{equation}
with $a_{s}=(\sqrt{\pi}\sigma_{s})^{-3}$ and $a_{p}=(\sqrt{\pi}\sigma_{p})^{-3}$.
The widths $\sigma_{s}$ and $\sigma_{p}$ are kept constant, $\sigma_{s}$
is fixed normalizing the space Gaussian to unity and $\sigma_{p}$
is correlated making use of the uncertainty relation to
$\sigma_{s}\cdot \sigma_{p}=\hbar/2$. In Eq.~(\ref{gauss-x}),
\begin{equation}
R^{\mu}_{i}(x) = \left( x^{\mu}-x_{i}^{\mu}(\tau) \right) -
		 \left( x_{\nu}-x_{i\nu}(\tau) \right)
		 u_{i}^{\mu}(\tau) u_{i}^{\nu}(\tau)
\label{dist-x}
\end{equation}
is the projection of the distance $x-x_{i}(\tau)$ on the hyperplane
perpendicular to the velocity $u_{i}(\tau)$. $\tau$ refers to the
eigentime of the particle.

A non-linear QHD model is adopted for the relativistic mean
field potential, with isoscalar, scalar $\sigma$- and vector
$\omega$-meson fields and with inclusion of the isovector channel
through the exchange of the virtual charged , scalar $\delta$- and,
vector $\rho$-mesons \cite{ref16,ref19,ref20,ref21}. In this model, in the mean field
approximation, the self energies $\Sigma_{s}$ and $\Sigma^{\mu}$
are proportional to the expectation values of the isoscalar and
isovector fields with coupling constants $g_{\sigma}$, $g_{\omega}$,
$g_{\rho}$ and $g_{\delta}$. The scalar and vector components are
given by
\begin{eqnarray}
\Sigma_{i}^{\mu} & = & g_{\omega}\omega^{\mu}(x)\pm g_{\rho}b^{\mu}(x)
\label{Sigma_v}\\
\Sigma_{si} & = & g_{\sigma}\sigma (x)\pm g_{\delta}\delta (x)
\label{Sigma_s}
\end{eqnarray}
with  the upper (lower) sign corresponds to the proton $p$ (neutron $n$).
The self energies characterize the in-medium properties of the nucleons
inside the hadronic environment in terms of kinetic momenta and
effective masses
\begin{eqnarray}
p_{i}^{*\mu} & = & p_{i}^{\mu} - \Sigma_{i}^{\mu}
\label{eff_mom}\\
m_{i}^{*} & = & M-\Sigma_{si}
\label{eff_mass}
\end{eqnarray}
which is different for protons and neutrons due to the appearance of the
isovector $\rho$ and $\delta$ mesons.

In the local density approximation, the scalar and vector meson fields,
determined by the scalar density $\rho_{s}$ and the baryonic
current $J_{\mu}$, respectively, result from the solution of
the corresponding equations
\begin{eqnarray}
m_{\sigma}^{2}\sigma (x) + a\sigma^{2}(x)+b\sigma^{3}(x) & = &
g_{\sigma}\rho_{s}(x)=g_{\sigma}\int d^{4}p^{*} \frac{m^{*}(x)}{E^{*}(x)}f(x,p^{*})
\label{sigma_eq}\\
\omega_{\mu}(x) & = & \frac{g_{\omega}}{m_{\omega}^{2}}J_{\mu}(x)
= \frac{g_{\omega}}{m_{\omega}^{2}}
\int d^{4}p^{*} p^{*}_{\mu} f(x,p^{*})
\label{omega_eq}\\
b_{\mu}(x) & = & \frac{g_{\rho}}{4m_{\rho}^{2}}J_{3\mu}(x)
= \frac{g_{\rho}}{4m_{\rho}^{2}}
\int d^{4}p^{*} p^{*}_{\mu} f_{3}(x,p^{*})
\label{rho_eq}\\
\delta_{\mu}(x) & = & \frac{g_{\delta}}{4m_{\delta}^{2}}\rho_{s3}(x)
= \frac{g_{\delta}}{4m_{\delta}^{2}}
\int d^{4}p^{*} \frac{m^{*}(x)}{E^{*}(x)} f_{3}(x,p^{*})
\label{delta_eq}\\
\quad .
\end{eqnarray}
The isospin vector phase-space and scalar densities are given
by $f_{3}=f_{p}-f_{n}$ and $\rho_{s3}=\rho_{sp}-\rho_{sn}$, respectively.

The equation of motion for the test particles are obtained by substituting
Eq.~(\ref{fs_1}) into Eq.~(\ref{rbuu}) and putting the collision term
equal to zero, such that
\begin{eqnarray}
\frac{d}{d\tau}x_i^{\mu}(\tau) & = & u_i^{\mu}
\nonumber\\
\frac{d}{d\tau}u_i^{\mu}
& = &
\frac{1}{m^{\ast}(x_{i})}
\sum_{j=1}^{A \cdot N_{test}} \frac{2}{\sigma_{s}^{2}}
\left[
	\frac{g_{\omega}^{2}}{m_{\omega}^{2}}
	u_{i\nu}
	\left(
		R_{j}^{\mu}(x_{i})u_{j}^{\nu}-R_{j}^{\nu}(x_{i})u_{j}^{\mu}
	\right)
\right.
\nonumber\\
	& - &
	\left.
	g_{\sigma}\frac{\partial\sigma (x_{i})}{\partial\rho_{s}}
	\left(
		R_{j}^{\mu}(x_{i})-u_{i}^{\mu}u_{i}^{\nu}R_{j\nu}(x_{i})
	\right)
\right]
\frac{\alpha_{s}exp(R_{j}^{2}(x_{i})/\sigma_{s}^{2})}{N_{test}}
\nonumber\\
& \pm &
\frac{1}{m^{\ast}(x_{i})}
\sum_{j=1}^{Z \cdot N_{test}} \frac{2}{\sigma_{s}^{2}}
\left[
	\frac{g_{\rho}^{2}}{4m_{\rho}^{2}}
	u_{i\nu}
	\left(
		R_{j}^{\mu}(x_{i})u_{j}^{\nu}-R_{j}^{\nu}(x_{i})u_{j}^{\mu}
	\right)
\right.
\nonumber\\
	& - &
	\left.
	\frac{g_{\delta}^{2}}{4m_{\delta}^{2}}
	\left(
		R_{j}^{\mu}(x_{i})-u_{i}^{\mu}u_{i}^{\nu}R_{j\nu}(x_{i})
	\right)
\right]	
\frac{\alpha_{s}exp(R_{j}^{2}(x_{i})/\sigma_{s}^{2})}{N_{test}}
\nonumber\\
& \mp &
\frac{1}{m^{\ast}(x_{i})}
\sum_{j=Z\cdot N+1}^{A \cdot N_{test}} \frac{2}{\sigma_{s}^{2}}
\left[
	\frac{g_{\rho}^{2}}{4m_{\rho}^{2}}
	u_{i\nu}
	\left(
		R_{j}^{\mu}(x_{i})u_{j}^{\nu}-R_{j}^{\nu}(x_{i})u_{j}^{\mu}
	\right)
\right.
\nonumber\\
	& - &
	\left.
	\frac{g_{\delta}^{2}}{4m_{\delta}^{2}}
	\left(
		R_{j}^{\mu}(x_{i})-u_{i}^{\mu}u_{i}^{\nu}R_{j\nu}(x_{i})
	\right)
\right]	
\frac{\alpha_{s}exp(R_{j}^{2}(x_{i})/\sigma_{s}^{2})}{N_{test}}
\label{rbuu2}
\quad .
\end{eqnarray}
Eq. (\ref{rbuu2}) is the Vlasov term of the transport equation in terms of the test
particle represention of the phase-space distribution function. It describes the
propagation of the test particles under the influence of the nuclear mean field,
which enters here via the coupling functions
$f_{n}=\frac{g_{n}^{2}}{m_{n}^{2}}$ ($n=\sigma,\omega$ and $n=\rho,\delta$,
for the isoscalar and isovector parts of the nuclear potential, respectively).
For charged baryons the Coulomb force represents an additional term in the
equation of motion.

In the RBUU code,  not only protons and neutrons are propagating separately
according to their hadronic fields and Coulomb interaction but also
$\Delta^{0,\pm,++}$-resonances. $N^{*}$-resonances are not fully accounted
for, which may set the upper limit of the incident energy of the code
to 1.5 GeV for the nucleon induced reactions. The self energies of the
resonances are built as \cite{ref27,ref28,ref29}
\begin{eqnarray}
\Sigma_{i}(\Delta^{-}) & = & \Sigma_{i}(n)
\nonumber\\
\Sigma_{i}(\Delta^{0}) & = & \frac{2}{3}\Sigma_{i}(n)+\frac{1}{3}\Sigma_{i}(p)
\nonumber\\
\Sigma_{i}(\Delta^{+}) & = & \frac{1}{3}\Sigma_{i}(n)+\frac{2}{3}\Sigma_{i}(p)
\nonumber\\
\Sigma_{i}(\Delta^{++}) & = & \Sigma_{i}(p)
\label{Sigma_Delta}
\quad ,
\end{eqnarray}
where $i=scalar, vector$.

Differently from $\Delta$ and nucleons, pions are propagated as
free particles with respect to the nuclear mean field, but experience
the Coulomb potential generated by all the other charged particles.

The collision integral, i.e. the r.h.s of Eq.~(\ref{rbuu}) is modeled by a
parallel ensemble Monte Carlo algorithm
\cite{ref13,ref14,ref16,ref28}. A geometrical
interpretation of the cross section is used, so that in each ensemble
two test-particles are allowed to collide if their relative distance
$d$ fulfills the relation:
\begin{equation}
d \leq \sqrt{\frac{\sigma_{tot}}{\pi N_{test}}}
\label{coll_dist}
\quad .
\end{equation}
The collision is accepted if the final state Pauli blocking is not violated.

The total cross section $\sigma_{tot}$ depends on the center of mass energy
and on the species of the incoming particles.  The distance $d$ is
calculated using the covariant calculation proposed by Kodama et al. \cite{ref30}.

The RBUU code describes the propagation and mutual interactions of nucleons,
Delta resonances and also $\pi$-mesons. The following hadronic reactions
are treated explicitly:
\begin{itemize}
\item $NN \to NN$ (elastic scattering)
\item $\Delta\Delta \to \Delta\Delta$ (elastic scattering)
\item $NN \leftrightarrow N\Delta$
\item $N\pi \leftrightarrow \Delta$
\end{itemize}
The different $NN \leftrightarrow N\Delta$ and $NN \leftrightarrow \Delta\Delta$
isospin channels are treated according to \cite{ref31}, in which the cross sections
are evaluated on the basis of the One Boson Exchange model and then fits to
the analytical calculations are given. For the elastic cross sections and the
angular distributions, the free parametrization according to Cugnon et al. \cite{ref33}
is used. Besides production and absorption in $NN$-channels, resonance states
can be populated also through $\pi N$ collisions. It should be mentioned that,
in contrast to other transport approaches \cite{ref7,ref9,ref10}, the RBUU code accounts
for both the relativistic kinematics and the isospin effects in resonance
production and decay,  due to the Lorentz structure of the isovector channel.

The mean field  is determined by four coupling constants,
$f_{i}=g_{i}^{2}/m_{i}^{2}$, $i=\sigma,~\omega,~\rho,~\delta$
and the two parameters of the non-linear self-interactions of the
$\sigma$-meson field $a,b$. In this work, for the isoscalar part
the parameter set NL of Ref. \cite{ref32} has been used.
For symmetric matter it gives reasonable values for the incompressibility
$K=200$ MeV with a nucleon effective mass of $m^{*}=0.83M$ at saturation
density $\rho_{0}=0.148$ $fm^{-3}$. In order to unravel the role played by
the different density behavior of the symmetry energy, we will discuss the
results obtained with the two parameter sets $NL\rho$ and $NL\rho\delta$
in the isovector channel. In $NL\rho$ and $NL\rho\delta$ models the
symmetry energy parameter is fitted at saturation to the value of $30.7$ MeV.
The values of all  the coupling constants are given in Table 1. In Fig.~\ref{Fig1}
we plot the density dependence of the symmetry energy for the $NL$, $NL\rho$
and $NL\rho\delta$  models. We note two features that will be relevant
for spallation reactions: i) The symmetry energy is in general larger
when isovector bosons are inserted (in the NL model only the kinetic
Fermi contribution is present); ii) The contribution of the $\delta$-meson
has different effects on the symmetry term, a decrease below saturation
and an increase above $\rho_{0}$ \cite{ref16,ref19,ref20}.

\nopagebreak
\begin{table}[t]
\begin{center}
\begin{tabular}{|c|c|c|c|c|c|c|}
\hline\hline
       & $f_{\sigma}~(fm^{-2})$   & $f_{\omega}~(fm^{-2})$ & $f_{\rho}~(fm^{-2})$     & $f_{\delta}~(fm^{-2})$     & $B~(fm^{-1})$ & $C$ \\
\hline\hline
   $NL$ &     9.3        &     3.6     & 0.0 & 0.0 &    0.015      &  -0.004  \\
\hline
   $NL\rho$ &     9.3        &     3.6    & 1.22  & 0.0 &    0.015      &  -0.004  \\
\hline
   $NL\rho\delta$ & 9.3        &     3.6    & 3.4  & 2.4 &    0.015      &  -0.004  \\
\hline\hline
\end{tabular}
\end{center}
\caption{{\it Coupling constants $f_{i}=\frac{g_{i}^{2}}{m_{i}^{2}}$
for $i=\sigma,~\omega,~\rho,~\delta$, $B=\frac{a}{g_{s}^{3}}$ and
$C=\frac{b}{g_{s}^{4}}$ for the different $NL-RMF$ models.}}
\label{table_1}
\end{table}

 It should be noted that the $NL$, $NL\rho$  and $NL\rho\delta$ parameters
 have been already used for pion/kaon production and isospin tracer
 calculations in relativistic heavy ion collisions with an overall
 agreement with the data \cite{ref13,ref14,ref28,ref29}.\\

 {\bf {\it Event Characterization}}\\

Due to the stochastic nature of the RBUU simulations the transport code is used
as an event generator.  In order to reduce statistical fluctuations 500 events
are analyzed  for each initial condition (target, beam energy and centrality).

An important issue for the description of spallation  data within the RBUU model
is an appropriate initialization of the ground state nuclei.  We will follow
here in particular the initialization of Ref. \cite{ref31}, which aims to construct the
static properties of the ground nuclear state such as binding energy and root
mean square radius. Moreover the evolution of these quantities should be stable
enough, and the behavior of the density and momentum distributions should be
approximately correct. To get the initial nuclear phase space distribution,
we first distribute the $N_{test}\times A$ test-particles according to a Saxon-Wood
shape with neutron (proton) radius $R_{n,p}(A)$ and diffuseness parameter $a_{n,p}(A)$
depending on the target mass $A$. The parameterizations of $R_{n,p}(A)$ and $a_{n,p}(A)$
as a function of mass $A$ for stable nuclei  have been deduced from the adjustment
of the Saxon-Wood function on the density matter distributions calculated with the model
of Ref. \cite{ref13}. The Saxon-Wood tail is cut off at $R_{max}=R_{n,p}+2a_{n,p}$. The next step is
to determine the local density of all  nucleons generated by all the other nucleons.
The initial momenta of $N_{test}\times A$ test-particles are randomly chosen between zero
and the local Fermi momentum: $p_{F}^{max}=\hbar c (3\pi^{2}\rho_{Sax})$, with $\rho_{Sax}$
being the corresponding Saxon-Wood -neutron or proton density. Finally, an initial
ground state configuration is accepted when (a) the sum of the total energy is equal
to $E_{B}/A\pm 0.5 MeV/A$, where $E_{B}/A$ is the ground state energy of a nucleus of
mass $A$ and charge $Z$ given by the liquid drop model and (b) the binding energy and
root mean square radius keep smooth variation with time and without spurious particle
emission.

The RBUU calculation is carried out up to a time scale referred to as the transition
time $t_{tr}$, when an excited compound system is formed, the "prefragment" source
that will decay via statistical emissions. At $t_{tr}$, the position of each nucleon
is used to calculate the distribution of mass and charge. The minimum spanning tree
method \cite{ref35} is employed, a prefragment is formed if the centroid distances are lower
than $R_{clus}$. In this paper, $R_{clus}$ is fixed at 4 fm. The prefragment thus
identified is then Lorentz boosted into his rest frame to evaluate the excitation
energies. The excitation energy ($\epsilon_{exc}$) is calculated as the difference
between the binding energy of the hot prefragment and the binding energy of this
prefragment in the ground state, evaluated as indicated before. When the prefragment
is in the excited state, the SM model \cite{ref23} is employed as an afterburner, which
is shown to be well suited for the description of the slow evaporated neutrons
\cite{ref11,ref23,ref34}.

In order to determine $t_{tr}$, it must be verified, whether information obtained
from the transport code are sensitive to the time duration of the first stage of the
reaction. For this purpose, the time variation of the average values (over 100 events)
of three physical quantities: excitation energy per nucleon $<\epsilon_{exc}>$,
angular momentum $<L>$ and mass number of the excited prefragments $<A_{exc}>$ after
the first RBUU stage, have been analyzed. As an example, Fig.~\ref{Fig2} shows the time evolution
of these quantities for $p+Pb$ collisions at 1.2 GeV, obtained with a $NL\rho$ interaction.
As one can see, large prefragments are produced at earlier times $5<t_{tr}<10~fm/c$
with a maximal value of $<\epsilon_{exc}>~MeV/A$ and relatively large $<L>$.
Then, both quantities $<\epsilon_{exc}>$ and $<L>$ drop very rapidly in the time
range 10-35 fm/c, in correspondence of a fast, pre-equilibrium, nucleon emission as
shown by the decrease of $<A_{exc}>$ in the bottom panel. The dropping of
$<\epsilon_{exc}>$, $<L>$ and $<A_{exc}>$ stops at around 40 fm/c and the average
values remain stable. At later times, $t>80~fm/c$, spurious increase (decrease)
of $<\epsilon_{exc}>$ ($<L>$ and $<A_{exc}>$) is observed, which is unphysical
and arises due to spurious numerical fluctuations. This may indicate that the
first dynamical stage of the reaction evolution should be terminated at about
80 fm/c and a statistical decay procedure could be inserted to get the final
reaction products.

Let us now check the dependence on the $t_{tr}$ choice of the final results.
In Fig.~\ref{Fig3} we show the neutron double differential cross section for the
reaction $p+Pb$ at 1.2 GeV. The line histograms denote the RBUU+SM calculations
using $NL\rho$ parameterizations with three different transition times, 60 fm/c
(thin lines), 75 fm/c (thick lines), and 100 fm/c (dot-dashed lines). As one can see,
calculations with different transition times yield the same results in the
low energy neutron region $\approx E_{n}\leq 30~MeV$. This may indicate that the
decay processes of the excited prefragments described by RBUU+SM are equivalent in
a $t_{tr}$ choice from 60 to 100 fm/c. This is important for the study of the
symmetry term dependence of this component of the neutron spectrum presented in the next section.

On the other hand, in the "cascade region" (see later), $30<E_{n}<60~MeV$, a
systematic difference between the calculations is seen as the angle increases.
This difference decreases within the time interval from 75 to 100 fm/c.  We
observe that the neutron data, in this energy range, are better reproduced by
the RBUU+SM at $t_{tr}=75~fm/c$. We, thus, use this value for all systems in
the present study, which is also supported by Fig.~\ref{Fig2}.

It should be pointed out here that in the SM model, standard symmetry coefficients
$\gamma_{sym}=23$ to 25 MeV of the fragment binding energy are used to describe 
light and heavy fragments produced in proton/nucleus-nucleus collisions \cite{ref24}. 
On the other hand, the symmetry energies within the different parameterizations used in 
the transport model may differ from those used in the statistical approach for 
densities beyond saturation. Therefore, it is worthwhile to examine the consequences 
of changing $\gamma_{sym}$ in SM calculations, as has been done in different studies 
that focused on isotopic distributions \cite{ref36_new}-\cite{ref39_new}. 
Fig. \ref{Fig3a} displays the evaporation neutron spectra obtained using RBUU (with NL)+SM calculations for p+Pb reactions at 0.8 GeV, and the symmetry term $\gamma_{sym}$ was varied between 5 to 25 MeV. As one can see, with $\gamma_{sym}=5$ MeV, the SM processes causes a slight increase of low energy neutrons at $E_{n}<10$ MeV. For larger values
of $\gamma_{sym}$, however, the evaporation neutron spectra are not much affected by the symmetry coefficient difference in the SM code. We note that at variance with the Heavy Ion Collisions studies of Refs.[36-38] in our case we do not have large compression-expansion effects. The final equilibrated source is then at densities close to the
saturation value. This has been checked analyzing the time evolution of
the root mean square radius of the composite system in the (p+Pb) case
at 0.8 GeV. After the thermal expansion the system shows a rather good stability with
smooth monopole oscillations, see also Ref. \cite{ref40_new}. This may imply that the
equilibrating source is formed at about normal nuclear density and it is not necessary to
use much reduced symmetry coefficients. Thus, in what follows the value of
$\gamma_{sym}=25$ MeV will be kept fixed in the SM code, which is supported by the
experimental analysis of $\gamma_{sym}$ with the SM model in $p,d$ and $\alpha$
induced reactions at relativistic energies up to 15 GeV  \cite{ref39_new}.


\section{Results and discussions}

In this section we present and discuss the predictions of the RBUU+SM model using
different NL parameter sets along with the  measurements \cite{ref36,ref37} of energy-angle
DDCS of neutrons induced by 0.8 GeV, 1.2 GeV and 1.6 GeV protons on $Fe$ and $Pb$.

The experimental energy spectra (see Figs.~\ref{Fig4} and \ref{Fig5} and similar ones) show
at $0^{o}$ two prominent peaks. These peaks are less pronounced at $10^{o}$ and are
insignificant at $25^{o}$ and larger. The (quasi-elastic) peak, characterized by a
narrow peak at a kinetic energy near that of the beam energy, is due to a
single ($p,n$) elastic scattering in the forward direction. The (quasi-inelastic) peak,
centered around $873~MeV$ and $760~MeV$ at $0^{o}$ and $10^{o}$, respectively, is
about $400~MeV$ wide and is thought to be due to $\Delta$-resonance excitation.
In addition to these two peaks, it seems that two components exist for all of the
spectra: one is a "shoulder" below $E_{n}\approx 20~(30)~MeV$ for Fe (Pb), the
other is a wide peak extending up to a few hundred MeV. The low energy neutrons mostly
come from evaporation of the excited target residues formed through the equilibration
process. The other component becomes less pronounced with increasing angles. This
component arises from "cascade" processes involving several $NN$-collisions.  Below we
are going to investigate the effect of the $NL$, $NL\rho$ and $NL\rho\delta$ models on
the three components of the neutron spectra by employing RBUU+SM code. We have performed
500 simulations at various impact parameters, from 0 to 4(7.5) fm for Fe(Pb),
respectively. In order to have sufficient statistics, calculations were done for angular
bins of $\pm 3.5$ at $0^{o}$ and $\pm 5$ for larger  angles.

Before confronting RBUU+SM with experimental  data, it is worthwhile to investigate
the sensitivity of the calculations to the number of test particles ($N_{test}$).
This has been done (not shown here) for $p+Fe$ reactions. We used $NL\rho$ parameter
set and the calculations were performed with 40 and 100 test particles per nucleon.
It was found that the overall behavior of the neutron DDCS are nearly the same for
both calculations at all angles. Since the computational time of the collision integral
scales as ($N_{test}\times A$) (parallel ensemble method), a value
of $N_{test}=40$ test particles per nucleon has been chosen.

All the results are presented and compared to data in the Figs.~\ref{Fig4}-\ref{Fig9}.
Before passing
to a more detailed analysis we like to note that all the components of the neutron
spectra are rather well reproduced by this relativistic transport approach. We remark
that no free parameters are used  while genuine relativistic effects are present via
the full covariance of the model and the natural presence of Lorentz forces due to
the coupling to vector fields.

\begin{figure}[t]
\begin{center}
\includegraphics[clip=true,width=0.75\columnwidth,angle=0.]
{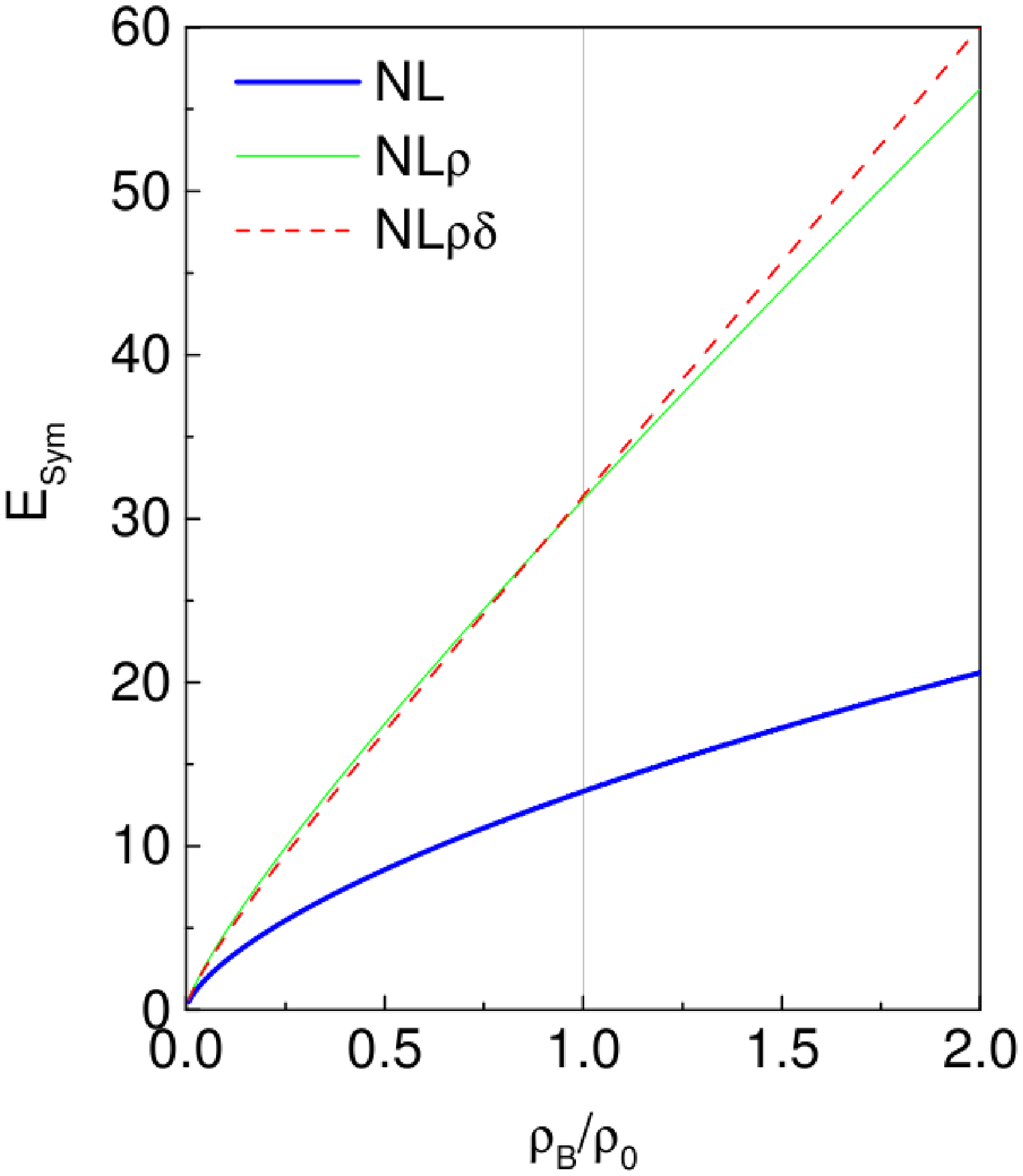}
\caption{\label{Fig1} 
(color online) Density dependence of the symmetry energy for the NL-RMF
models.
}
\end{center}
\end{figure}

\begin{figure}[t]
\begin{center}
\includegraphics[clip=true,width=0.75\columnwidth,angle=0.]
{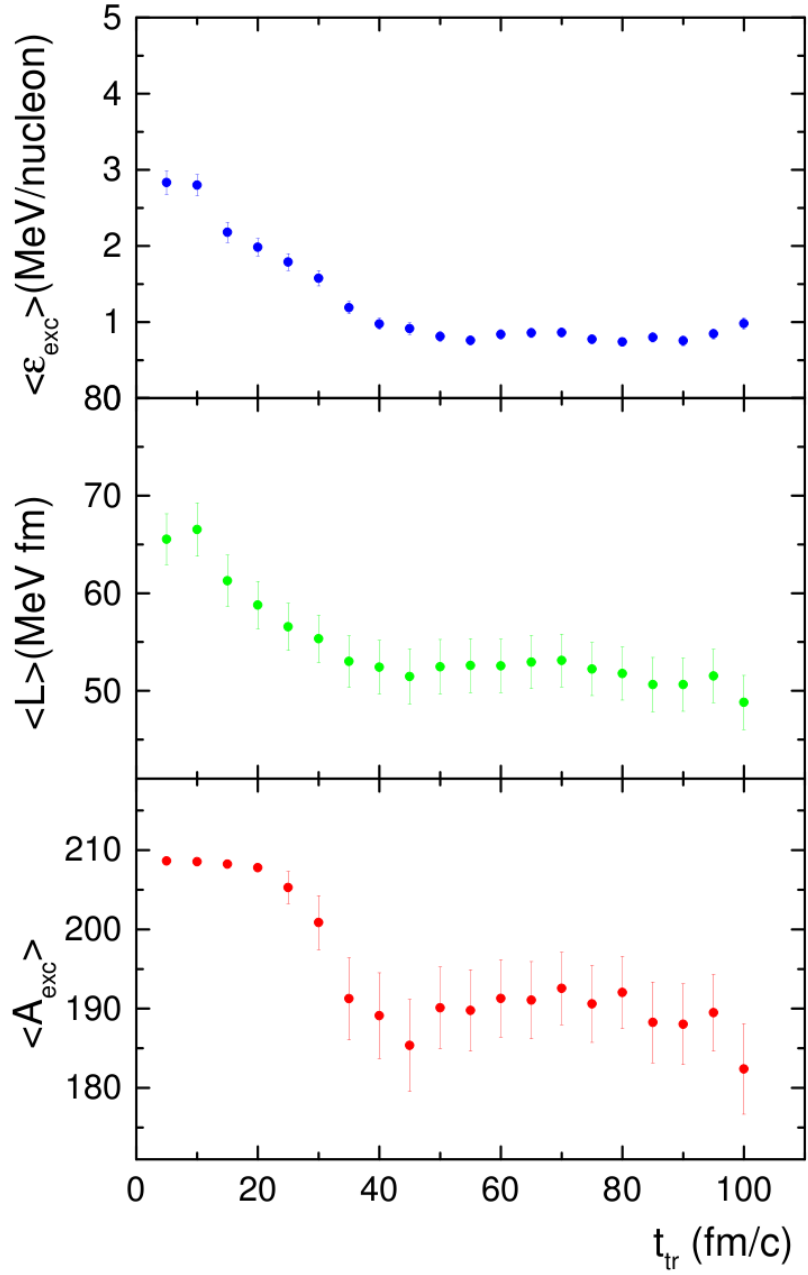}
\caption{\label{Fig2} 
(color online) Calculated average excitation energy ($<\epsilon_{exc}>$),
angular momentum ($<L>$) and mass number of
prefragments $<A_{exc}>$ as a function of transition
time ($t_{tr}$) after the RBUU (with $NL\rho$) initiated by a proton
on Pb at 1.2 GeV.
}
\end{center}
\end{figure}

\begin{figure}[t]
\begin{center}
\includegraphics[clip=true,width=0.75\columnwidth,angle=0.]
{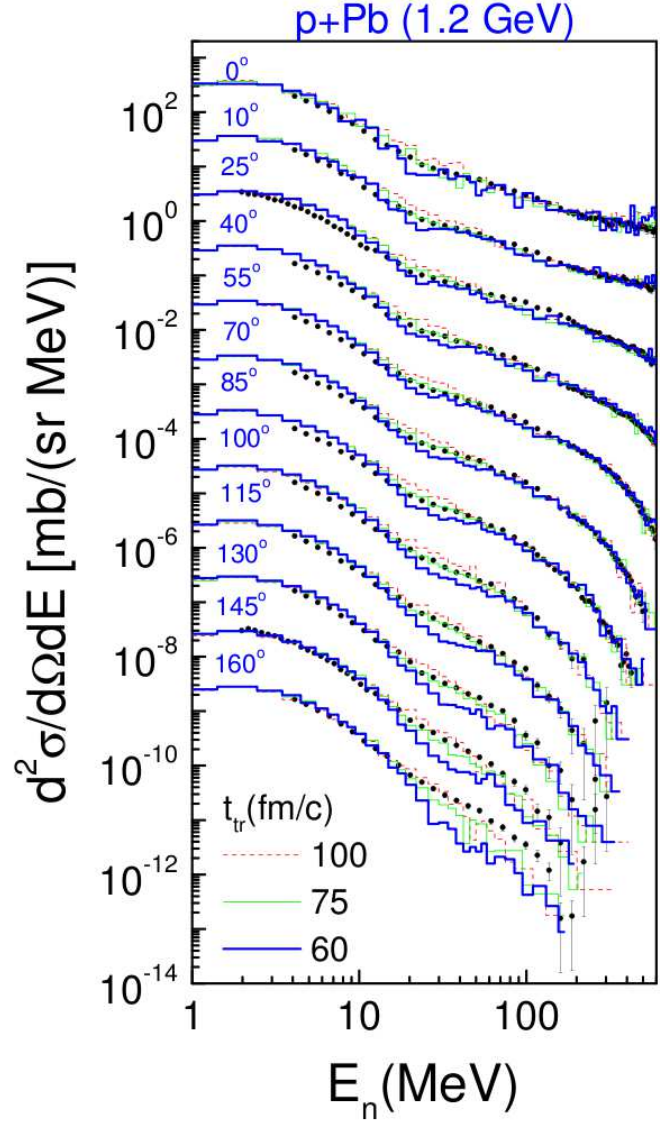}
\caption{\label{Fig3} 
(color online) Neutron energy-angle double differential cross sections for p+Pb
at 1.2 GeV calculated by RBUU(with $NL\rho$)+SM with different transition
times ($t_{tr}$). Data (solid circles with error bars) are from
\protect\cite{ref36}.
For clarity, only the histograms and the data for the smallest angles
are given in absolute value. The other ones have been multiplied
by $10^{-1},~10^{-2},~\ldots$ for other angles in increasing order.
}
\end{center}
\end{figure}

\begin{figure}[t]
\begin{center}
\includegraphics[clip=true,width=0.75\columnwidth,angle=0.]
{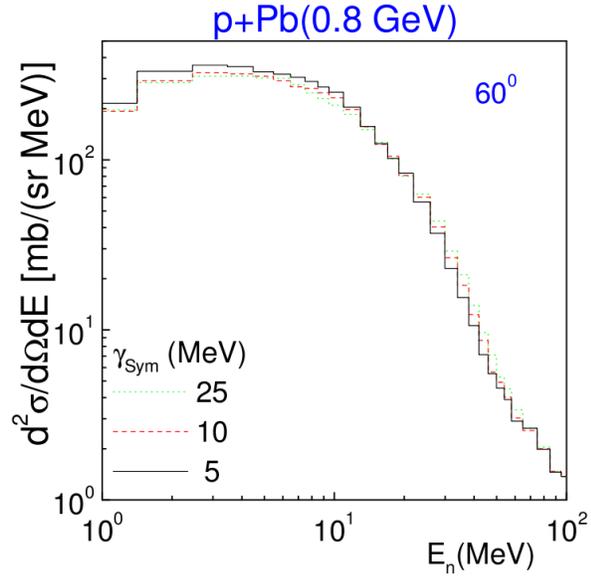}
\caption{\label{Fig3a} 
(color online) Evaporation neutron double differential cross sections for p+Pb reactions at 0.8 GeV calculated by RBUU(with NL)+SM. The lines are the results obtained by changing the symmetry coefficients ($\gamma_{sym}$) in the SM code.
}
\end{center}
\end{figure}

\begin{figure}[t]
\begin{center}
\includegraphics[clip=true,width=1\columnwidth,angle=0.]
{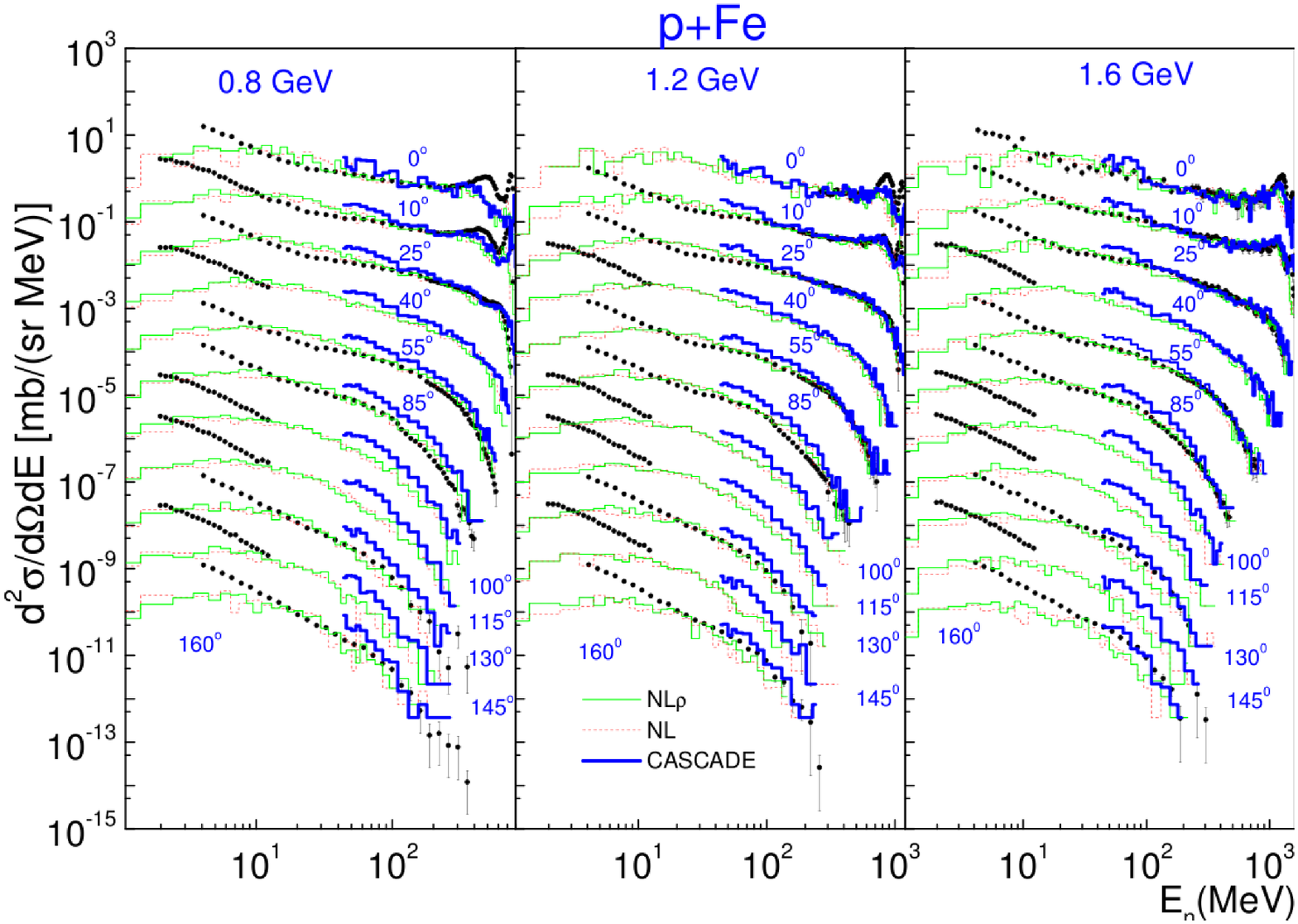}
\caption{\label{Fig4} 
(color online) Neutron energy-angle double differential cross sections
for p+Fe interactions at 0.8 GeV(left panel), 1.2 GeV(middle panel)
and 1.6 GeV (right panel) as compared to RBUU calculations
(without evaporation). The dashed and thin lines denote calculations
with $NL$ and $NL\rho$, respectively. The thick lines show calculations
without mean field. The data \protect\cite{ref36}
(solid circles with error bars)
and the histograms  for the smallest angles are given in absolute value.
The other ones have been multiplied by $10^{-1},~10^{-2},~\ldots$
for other angles in increasing order.
}
\end{center}
\end{figure}

\begin{figure}[t]
\begin{center}
\includegraphics[clip=true,width=1\columnwidth,angle=0.]
{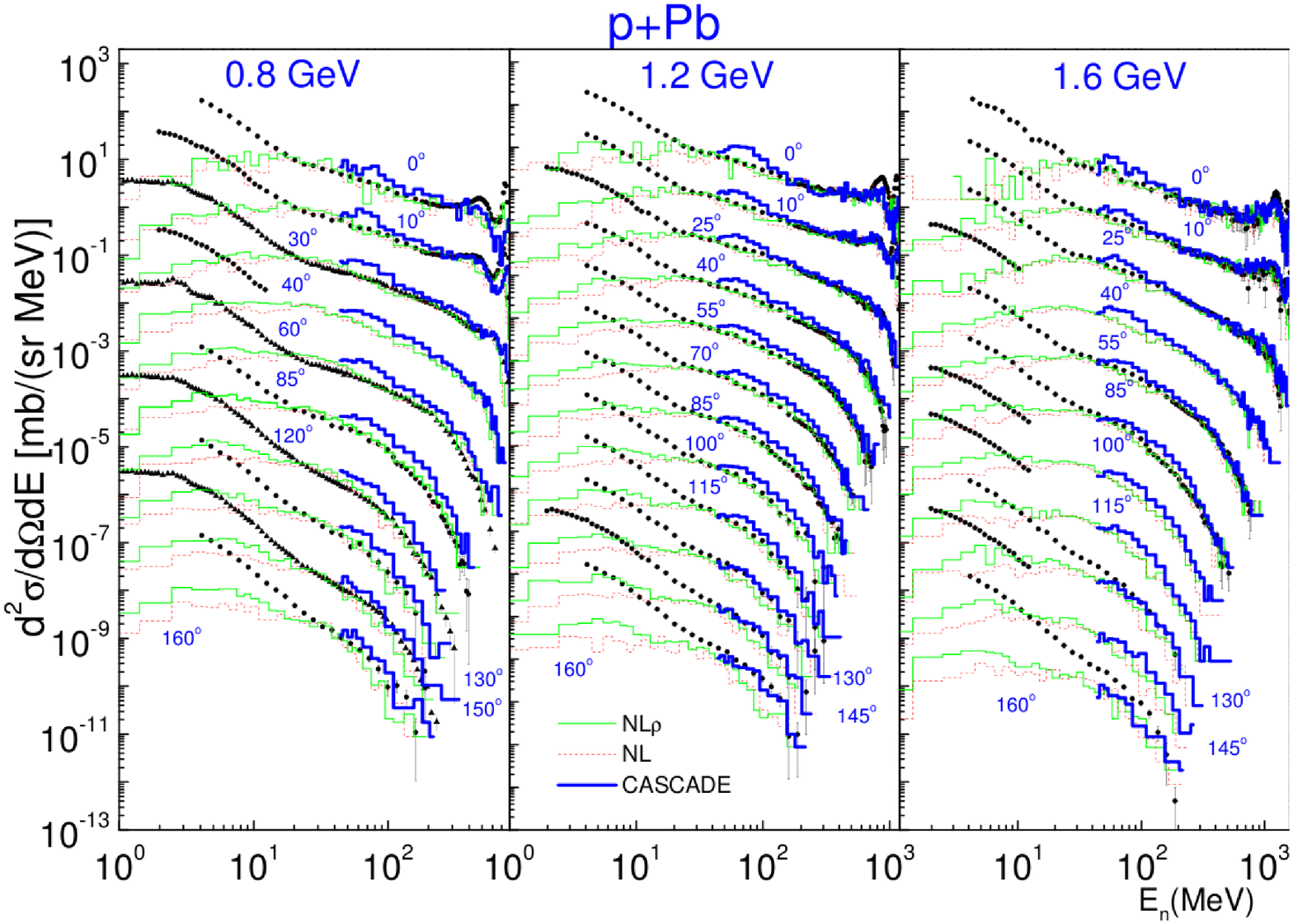}
\caption{\label{Fig5} 
(color online) Same as Fig.5, but for p+Pb reactions. Solid triangles
with error bars (left panel) are data taken from \protect\cite{ref37}.
}
\end{center}
\end{figure}

\begin{figure}[t]
\begin{center}
\includegraphics[clip=true,width=1\columnwidth,angle=0.]
{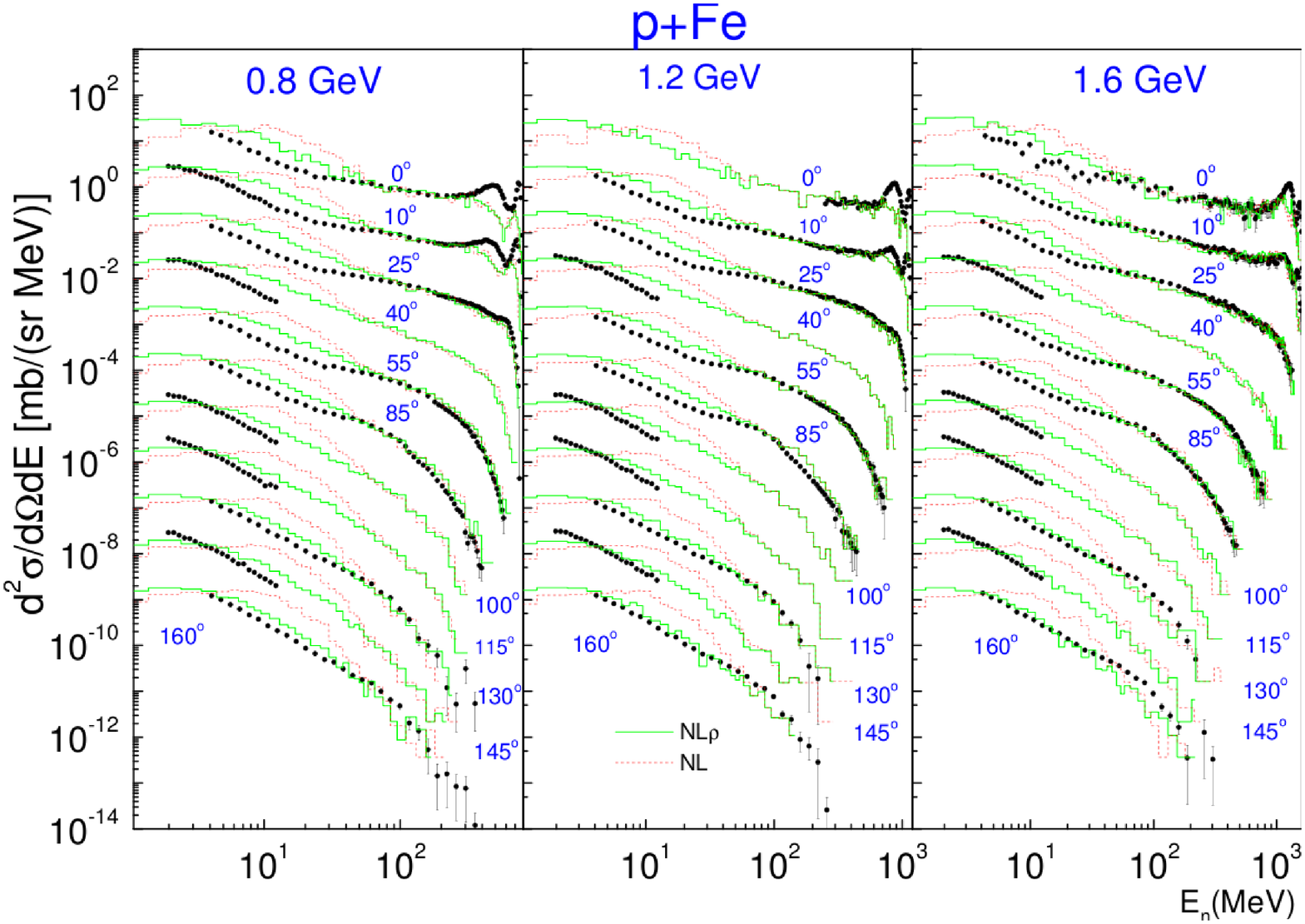}
\caption{\label{Fig6} 
(color online) Same as Fig.5, but the solid and dashed lines denote
the RBUU+SM with $NL\rho$ and $NL$, respectively.
}
\end{center}
\end{figure}

\begin{figure}[t]
\begin{center}
\includegraphics[clip=true,width=1\columnwidth,angle=0.]
{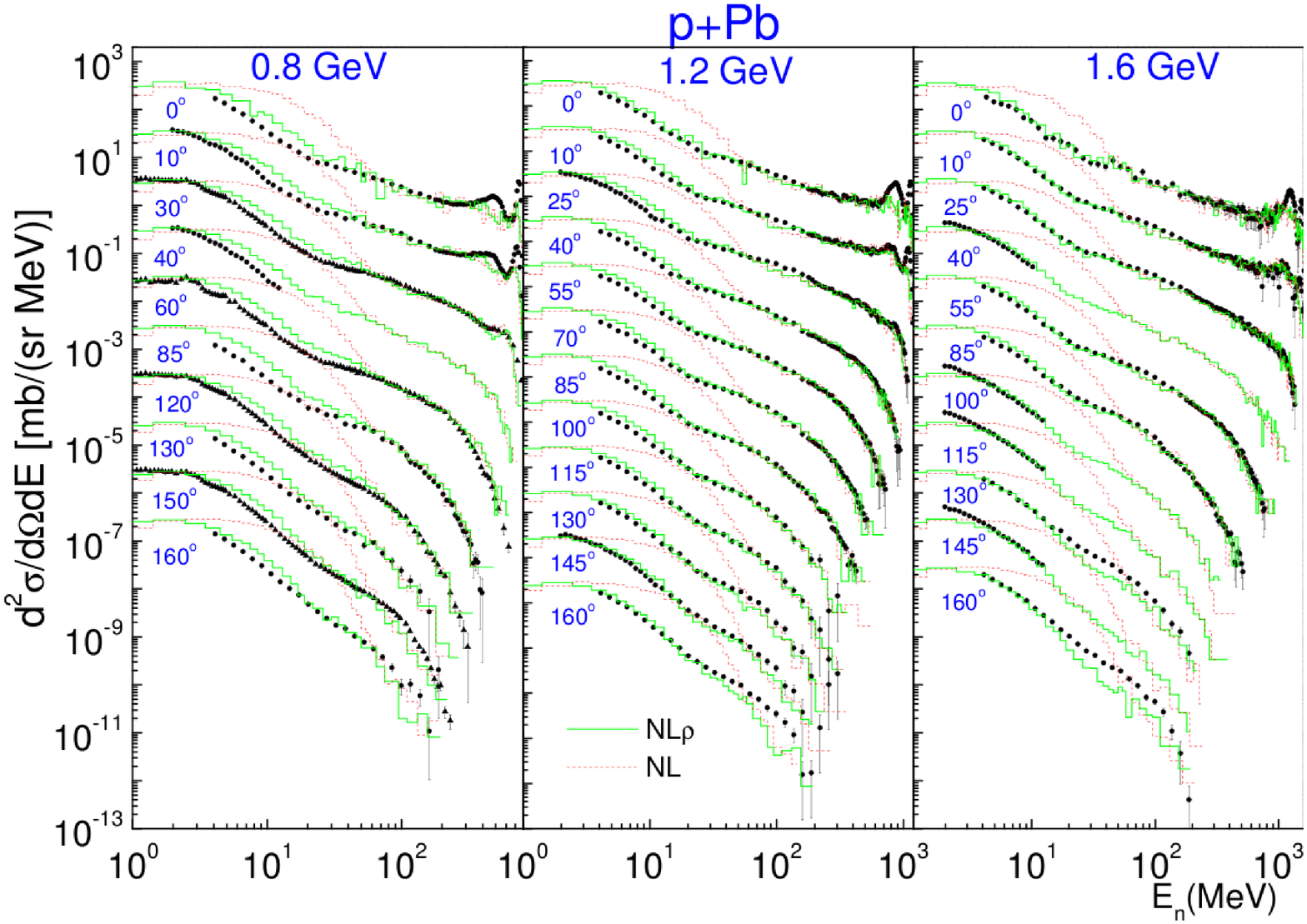}
\caption{\label{Fig7} 
(color online) Same as Fig.7, but for p+Pb reactions. Solid triangles
with error bars (left panel) are data taken from \protect\cite{ref37}.
}
\end{center}
\end{figure}

\begin{figure}[t]
\begin{center}
\includegraphics[clip=true,width=1\columnwidth,angle=0.]
{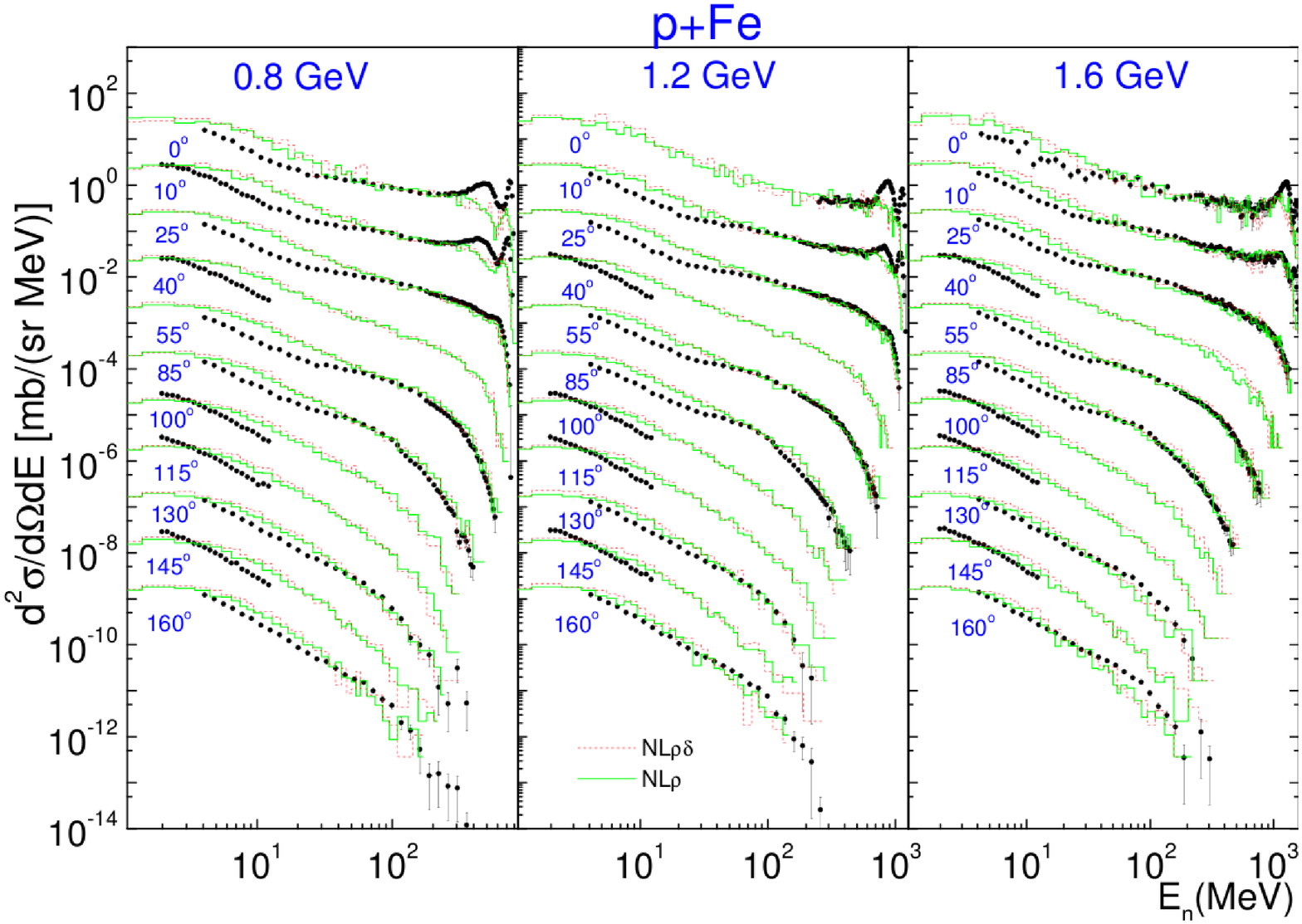}
\caption{\label{Fig8} 
(color online) Same as Fig.5, but the dashed and solid lines denote the
RBUU+SM results with $NL\rho\delta$ and $NL\rho$, respectively.
}
\end{center}
\end{figure}

\begin{figure}[t]
\begin{center}
\includegraphics[clip=true,width=1\columnwidth,angle=0.]
{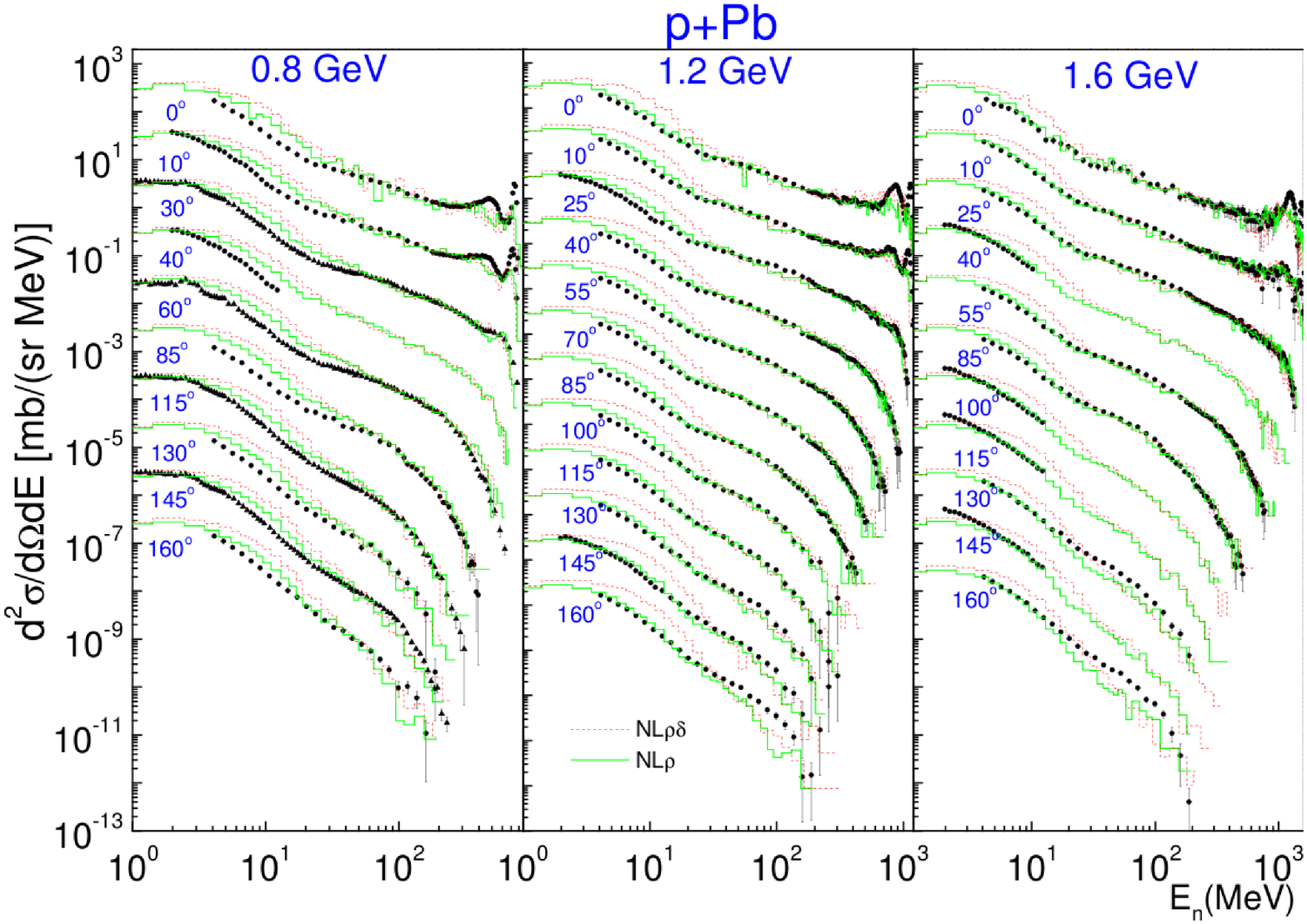}
\caption{\label{Fig9} 
(color online) Same as Fig.9, but for p+Pb reactions. Solid triangles
with error bars (left panel) are data taken from \protect\cite{ref37}.
}
\end{center}
\end{figure}

Let us first focus on the role of symmetry energy on the neutron spectra. This is
illustrated in Figs.~\ref{Fig4}-\ref{Fig7} by comparing the RBUU calculations
without (Figs.~\ref{Fig4} and \ref{Fig5})
and with (Figs.~\ref{Fig6} and \ref{Fig7}) evaporation using $NL$ (dot-dashed lines) and
$NL\rho$ (thin lines) models. The two models differ only in the isospin part, with
$\rho$-coupling giving an almost linear density dependence of the symmetry energy
with value $\approx 30.7~MeV$ at saturation (see Fig.~\ref{Fig1}).

Figs.~\ref{Fig4} and \ref{Fig5} show that the results of both calculations, $NL$ and $NL\rho$, are close
together and can describe the high energy part of the neutron spectra
(above $E_{n} \approx 20$ and $30~MeV$ for p+Fe and Pb, respectively). It
is interesting to see that without statistical decay the "low energy shoulder" is
absolutely missed.  For the more neutron-rich Pb target some difference originating
from the $\rho$-meson becomes more pronounced in the lower part of the neutron
energy spectrum (below 30 MeV), especially with increasing emission angle.
Such behaviour can be related to the difference in the symmetry energy: neutrons
experience a stronger repulsive field going from $NL$ to $NL\rho$, at any
density (see Fig.~\ref{Fig1}) \cite{ref16,ref20}.

In fact the largest isospin effects are observed just on the low energy
evaporation component which appears when the afterburner statistical decay
is included. In Figs.~\ref{Fig6} and \ref{Fig7} we compare the full RBUU+SM calculations
(with evaporation) along with the experimental neutron spectra. As one can see,
the $NL$ results, with no (potential) symmetry energy, show a broad maximum
at $E_{n}\approx 3-50~MeV$ in all angular intervals for all studied collisions.
On the other hand, the $NL\rho$ calculations reduce the neutron yield in this
energy region and more closely reproduce the experimental data, especially as
the incident energy increases. The importance of the isovector interaction is
shown to increase when both the incident energy and the isospin asymmetry of the
target system increase. It should be pointed out here that, the effect of the
symmetry potential on the three components of the neutron spectra has already
been discussed by a non-relativistic QMD approach for p+Pb at 1.2 and 1.6 GeV
\cite{ref23}. It was shown that of all the three components of the neutron spectra, the
evaporation part is the most sensitive to the symmetry potential.

Next we compare, in Figs.~\ref{Fig8} and \ref{Fig9}, the RBUU+SM results using $NL\rho$ without
(solid lines) and with (dot-dashed lines) the $\delta$-field. In the case of
p+Fe interactions, there is almost no significance differences between
$NL\rho$ and $NL\rho\delta$ results. However, for p+Pb interactions, with
increasing neutron to proton asymmetry, $NL\rho\delta$ generates an excess of
slow neutrons ($E_{n}\approx 3-50 MeV$) compared to $NL\rho$.

\begin{figure}[t]
\begin{center}
\includegraphics[clip=true,width=1\columnwidth,angle=0.]
{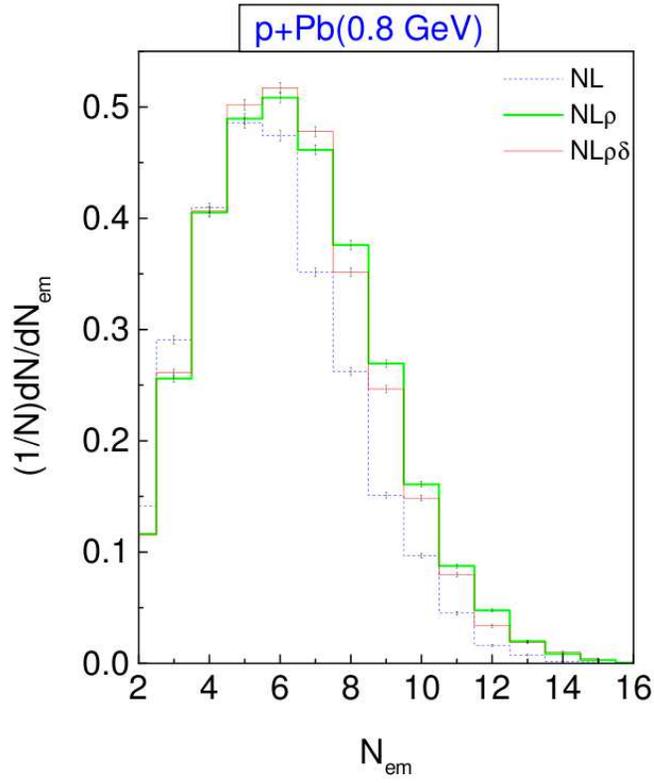}
\caption{\label{Fig10} 
(color online) Normalized distribution of pre-equilibrium neutrons
for p+Pb reactions at 0.8 GeV. The lines represent the
RBUU (without SM) with different NL-RMF models.
}
\end{center}
\end{figure}

The origin of the different behaviour between the results in the slow neutron
energy region (<50 MeV) is clearly due to the excitation energy and isospin
content of the hot prefragments. The properties of these prefragments at the
final stage of the interactions (at 75 fm/c) are affected by the number of
pre-equilibrium neutrons ($N_{em}$) emitted at the earlier times: less $N_{em}$
leads to larger excitation energy and larger N/Z of the prefragment. The
$NL\rho$ calculations of the p+Pb interactions at 1.2 GeV show, in Fig.~\ref{Fig2},
that nearly 20 nucleons are emitted at the earlier times (between 20 and 40 fm/c).
In Fig.~\ref{Fig10} we examine the distributions of  emitted neutrons $N_{em}$ before 75 fm/c
by employing RBUU with $NL$, $NL\rho$  and $NL\rho\delta$ cases for the p+Pb reactions
at 0.8 GeV. One observes that for $N_{em}\geq 8$, less pre-equilibrium neutrons are emitted
with $NL<NL\rho\delta <NL\rho$. Consequently, the excitation energy and
isospin asymmetry of the prefragments should appear larger for $NL$ calculations with respect
to $NL\rho$ (and $>NL\rho\delta$) as observed. The sequence ($NL<NL\rho\delta <NL\rho$)
agrees with the sequence of the three forms of the symmetry term at the densities
below saturation (see Fig.~\ref{Fig1}), which confirms that $N_{em}$ is generated at earlier times
where an expansion mode, due to thermal pressure, is likely occurred.

The sensitivity of the evaporation component of the neutron spectrum to the low
density behavior of the symmetry term can be then used to disentangle among
different effective fields and to get an independent information on the symmetry
energy below saturation, see some general discussions in the reviews
\cite{ref16,ref38}.
The best agreement with the data in the more n-rich Pb case is obtained within
the $NL\rho$ choice, which corresponds to a rather stiff increase of the symmetry
energy below saturation, roughly proportional to the total density. It is also
interesting to notice (see Figs.~\ref{Fig7} and \ref{Fig9}) that the $NL$ and $NL\rho\delta$ give similar
extra yields with respect to the $NL\rho$ case, which implies that the excitation
energies and N/Z of the Pb hot prefragments are similar. This can be also explained
by considering the weaker symmetry term at densities below saturation (see Fig.~\ref{Fig1})

\begin{figure}[t]
\begin{center}
\includegraphics[clip=true,width=0.75\columnwidth,angle=0.]
{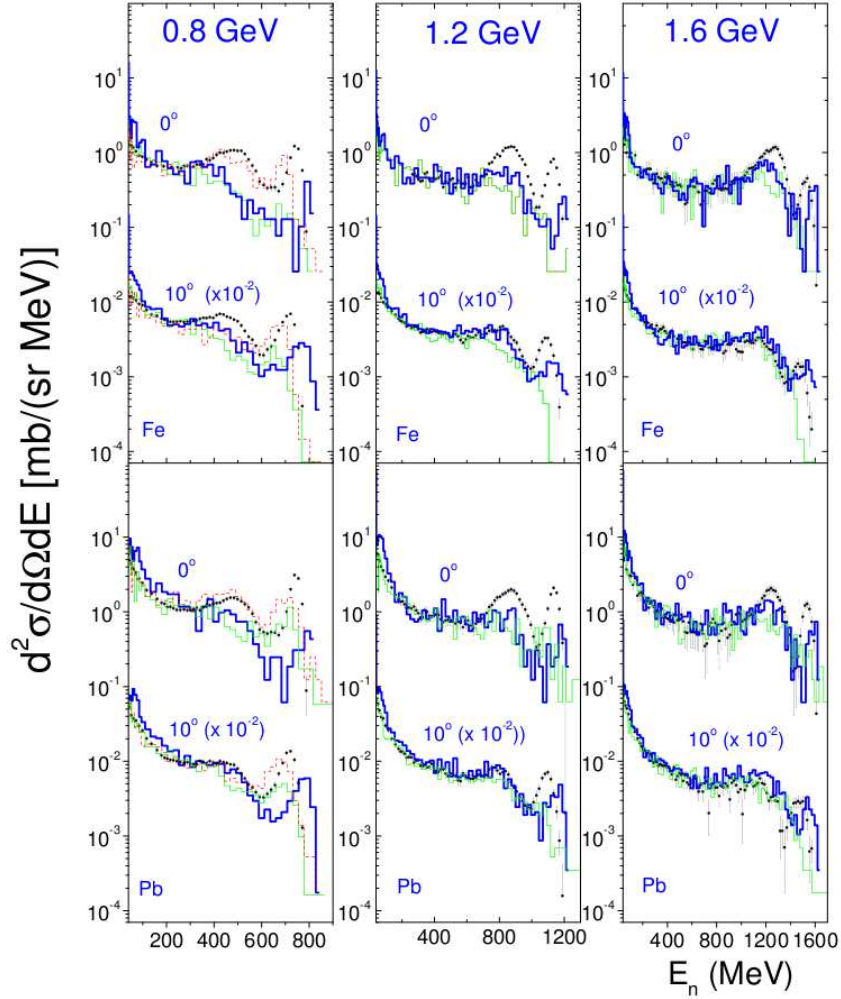}
\caption{\label{Fig11} 
(color online) Neutron energy spectrum at $0^{o}$ and $10^{o}$
from 0.8 GeV (left panel), 1.2 GeV (middle panel) and
1.6 GeV (right panel) proton interactions with Fe (top panels)
and Pb (bottom panels) targets. The thin lines denote
RBUU calculations with $NL\rho$, while the thick lines are those
without mean field. The dashed lines denote
RBUU (with $NL\rho$) calculations that include the medium modified
angular distributions of Ref. \protect\cite{ref40}.
}
\end{center}
\end{figure}

Finally, let us identify the effects of the mean field on the high energy part of
the neutron spectra ($E_{n}>40~MeV$) for the reactions under study. This is
illustrated in Figs.~\ref{Fig4} and \ref{Fig5} and Fig.\ref{Fig11} which show a close up of this region at
very forward angles ($\theta\leq 10^{o}$). Two different scenarios are explored:
RBUU calculations with $NL\rho$ (thin lines) are contrasted by RBUU with deactivated
mean field (CASCADE) (thick lines). We observe that at $10^{o}<\theta\leq 55^{o}$
the $NL\rho$ calculations are practically identical to CASCADE calculations as the
incident energy increases, showing the mean field effects to become negligible.
At $55^{o}<\theta\leq 160^{o}$ the $NL\rho$ calculations give lower values compared
to CASCADE calculations, in good agreement with the data.  However, the
$NL\rho$ calculations underestimate the high energy neutrons (70-110 MeV)
at $\theta=160^{o}$ for p+Fe at 1.6 GeV and p+Pb at 1.2 GeV, contrary to
CASCADE calculations. Note that, both calculations resemble each other
at $\theta=160^{o}$ for p+Pb at 1.6 GeV. Thus, it seems that the high energy
neutron spectra evaluated by RBUU (with and without $NL\rho$) have different beam
energy dependence compared to the experimental data. It should be pointed out that
our results in the backward direction are nevertheless better with our present
RBUU code than with the non-relativistic INC \cite{ref8}, QMD \cite{ref22} and
BUU \cite{ref39} calculations.

Fig.~\ref{Fig11} shows how the quasi elastic and quasi inelastic peaks are changed due to the
mean field. In the CASCADE calculations, the maximum of the calculated quasi-elastic
peak is shifted toward a higher energy in comparison to the data. The shape and
yield of the peaks are in better agreement with the CASCADE calculations at 1600 MeV
(and at $10^{o}$) than at 800 and 1200 MeV. On the other hand, as the mean field
calculations are taken into account in RBUU, the maximum outgoing neutron energy
is strongly attenuated at incident energy of 1200 MeV and 1600 MeV. As for the
quasi-inelastic peak, i.e., the peak located at the beam energy minus $\approx 300~MeV$,
the evolution of the shape  with the incident energy follows more or less the data
for both calculations. We notice that the $NL\rho$ calculations have a tendency to
become smaller compared to the CASCADE ones at lower incident energy and need
to be improved.

In \cite{ref40} the quasi-elastic and inelastic peaks at 1.2 GeV were underestimated
by the ultrarelativistic QMD (UrQMD) calculations, which was attributed to
the free angular distributions of the scattered NN-elastic and inelastic
collisions adopted in the UrQMD model, and it was suggested to introduce
medium modified angular distributions (MMAD) into the UrQMD model for reducing
the deviation. We then further test the influence of the MMAD on the reactions
under study. We introduce the MMAD adopted in \cite{ref40} in the collision term of
RBUU (with $NL\rho$) code. The dot-dashed lines in Fig.\ref{Fig11} illustrate these
calculations for p+Pb at 800 MeV.  As one can see, the introduction of MMAD
in RBUU improves the intensity of both the quasi-elastic and inelastic peaks.
For the location, however, the maximum of the calculated quasi-elastic
peak is 50 MeV downward shift in comparison to the data. At higher incident
energy (not shown here), this shift is increased even further to 100 and
150 MeV for  the reactions under study at 1200 MeV and 1600 MeV, respectively.
We attribute this failure to the linear energy dependence of the Schrodinger
equivalent optical potential of the NL-RMF models at high energies \cite{ref41}.
Indeed, Dirac phenomenology on elastic proton-nucleus scattering predicts
a nonlinear energy dependence of Schrodinger equivalent optical potential
for incident energy above 400 MeV, which leads to a softening of the optical
potential and, consequently,  may improve the dynamics of  NN-elastic
collisions at high energies.

\section{Summary and conclusions }

We have studied proton-induced Fe and Pb reactions at 0.8, 1.2 and 1.6 GeV
by a hybrid model, using the RBUU transport approach for the first
pre-equilibrium stage and the SM decay model for the second slow stage
of the reaction. The NL-RMF model of the present approach consists
of isoscalar ($\sigma - \omega$ mesons) and isovector ($\rho - \delta$ mesons)
parts with different Lorentz covariant properties. The impact of different
NL-RMF models ($NL$, $NL\rho$  and $NL\rho\delta$) on the neutron spectra
of the studied reactions has been investigated and the following conclusions
can be drawn:
\begin{itemize}
\item Some effects of the mean fields are found on the high energy part of the neutron spectra, when switching off the mean field and turning to the CASCADE mode within the RBUU calculations.
\item
The low energy neutron spectrum ( $E_{n}\approx 3-50~MeV$) is found to be quite sensitive
to the details of  the relativistic structure of the isovector part of the mean field:
the inclusion of the $\rho$-field improves  the comparison between theory and
experiment to large extent. The effect is related to the properties
(rate and isospin content) of the fast nucleon emission during the thermal
expansion of the system and so it is sensitive to the low density behaviour
of the symmetry term.
\item
The competition between a scalar attractive ($\delta$) and a vector
repulsive ($\rho$) fields affects low energy neutrons ($E_{n}\leq 30~MeV$)
only for the more neutron-rich Pb target.
\item
The introduction of medium modified angular distribution in RBUU
(with $NL\rho$) improves the intensity of p+Fe and Pb reactions at 0.8 GeV.
\end{itemize}

Thus, the present comparisons suggest a better agreement with $NL\rho$ model,
however, several other aspects of the mean field dynamics of the RBUU model
need further work. For example, the nucleon self-energies of  the current
version of the model have no momentum dependencies, which might improve
the extremity of the calculated spectra at $0^{o}$ and $160^{o}$.
This work is in progress.

\begin{center}
{\bf Acknowledgements}
\end{center}
Kh. A. and N. F. would like to thank the staff members of the Theory Group
of the  Laboratori Nazionali del Sud (LNS), INFN, in Catania for the
hospitality during their visit to LNS.



\newpage


\newpage


\newpage


\newpage


\newpage



\newpage


\newpage


\newpage


\newpage


\newpage


\newpage


\newpage


\end{document}